\documentclass[prb,twocolumn,showpacs]{revtex4}
\usepackage{graphicx}

\begin{document}
\newcommand{\icm}{\ensuremath{\mbox{cm}^{-1}}}
\newcommand{\asf}{$\rm (TMTTF)_2AsF_6$}
\newcommand{\pf}{$\rm (TMTSF)_2PF_6$}
\newcommand*{\augsburg}{Experimentalphysik\ II, Institut f\"ur Physik, Universit\"at Augsburg, D-86135 Augsburg, Germany}
\newcommand*{\stuttgart}{1.\ Physikalisches Institut, Universit\"at Stuttgart,
D-70550 Stuttgart, Germany}

\title{Pressure-induced deconfinement of the charge transport\\ in the quasi-one-dimensional
Mott insulator \asf}

\author{A. Pashkin} \altaffiliation[Present address: ]{\augsburg}
\author{M. Dressel}
\author{C. A. Kuntscher$^{*,}$} \email{christine.kuntscher@physik.uni-augsburg.de}

\affiliation{\stuttgart}

\begin{abstract}
We studied the pressure dependence  of the
room-temperature infrared reflectivity of \asf\ along all three
optical axes. This anisotropic organic compound consists of
molecular stacks with orbital overlap along the $a$ direction; due
to electronic correlations the system is a quasi-one-dimensional
Mott insulator with a charge gap $\Delta_{\rho}\approx 70$~meV.
The gap is gradually reduced with increasing external pressure,
accompanied by the onset of a Drude contribution along the
stacking direction. In the
perpendicular $b'$ direction a Drude-like optical response is
observed for pressures above 2~GPa. This behavior
is interpreted in terms of a deconfinement of the electrons in a
one-dimensional Mott insulator, i.e.\ an insulator-to-metal
transition which occurs when the interchain transfer integral
$t_b$ is approximately equal to half of the charge gap
energy. We estimate the values of $t_b$ and the Luttinger liquid
parameter $K_\rho$ as a function of pressure.
\end{abstract}


\maketitle

\section{Introduction}

\begin{figure}
  \includegraphics[width=1\columnwidth]{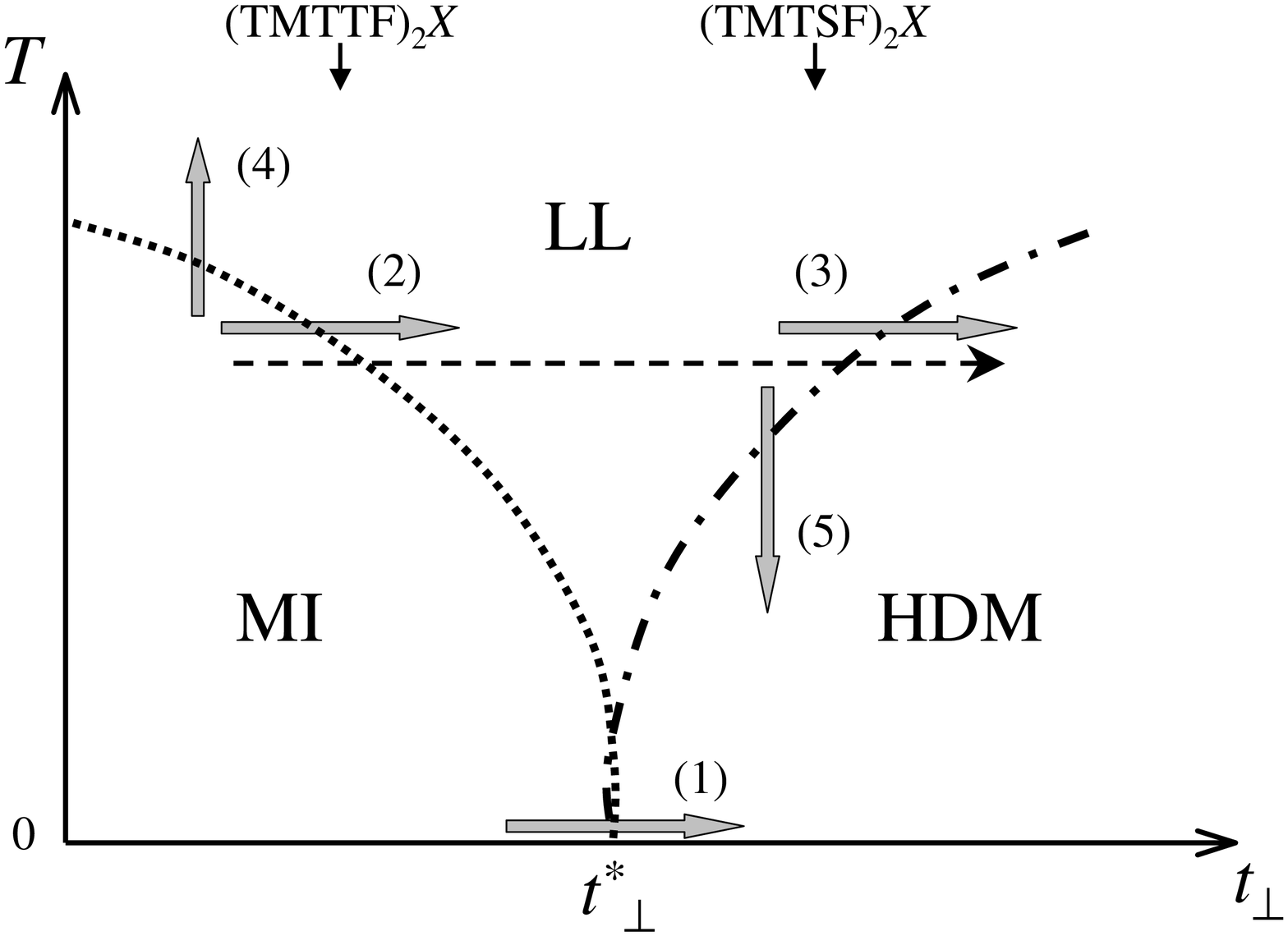}\\
  \caption{Schematic phase diagram of the deconfinement
  transition for a system of weakly coupled conducting chains according to Refs.\
  \onlinecite{Biermann01,Giamarchi04}. Transition
  from a Mott insulator (MI) to a high-dimensional metallic (HDM) state
  occurs at $T=0$ when $t_\perp$ reaches a critical value $t_\perp^*$ (trajectory 1).
  At high enough temperature, the increase of $t_\perp$ leads to a transition
  from a Mott insulating to a 1D Luttinger liquid (LL) conducting state (trajectory 2)
   and finally to a dimensional crossover into an HDM state (trajectory
   3). The temperature change at a fixed value $t_\perp$ may also lead to a
   transition from a Mott insulating to a LL state (trajectory 4), or to a dimensional crossover
   (trajectory 5).
   The horizontal dashed line indicates the path of our pressure study.\label{fig:deconfine}}
\end{figure}

The properties of quasi-one-dimensional conductors have been
subject of intensive studies for half a century. A theory based on
the one-dimensional (1D) Hubbard model predicts a non-Fermi-liquid
metallic state.\cite{Schulz91,Voit94,Giamarchi04} The situation,
however, develops differently when the filling of the system is
commensurate, since in this case, electronic interactions may
cause a Mott insulating state. Although this phenomenon is
possible in all dimensions, it is particularly strong in the 1D
case via umklapp process.\cite{Giamarchi97} The influence of the
interchain coupling on the electronic properties of a half- and
quarter-filled quasi-1D interacting electron system was studied
theoretically.\cite{Suzumura98,Tsuchiizu99a,Tsuchiizu99,Tsuchiizu01,LeHur01,Biermann01,Giamarchi04}
It is predicted\cite{Biermann01,Giamarchi04} that with increasing
interchain transfer integral $t_{\bot}$ such a system undergoes a
deconfinement transition into a metallic state. The schematic
phase diagram illustrating this phenomenon is shown in
Fig.~\ref{fig:deconfine}. Strictly speaking, the deconfinement
transition takes place at $T=0$ when $t_{\bot}$ becomes larger
than a certain value $t^{*}_{\bot}$ and can be referred to as a
quantum phase transition. For $t_{\bot} < t^{*}_{\bot}$ the system
is in a Mott insulating state and the single-particle interchain
hopping is completely suppressed. If the interchain transfer
integral exceeds a critical value $t^{*}_{\bot}$, the insulating
state is broken and single-particle interchain hopping becomes
possible. Thus, the system evolves from a 1D Mott-Hubbard
insulator to a high-dimensional metallic state (trajectory 1 in
Fig.~\ref{fig:deconfine}). The theoretical
study\cite{Suzumura98,Tsuchiizu99a} of two coupled Hubbard chains
predicts a close relation of the interchain transfer integral to
the charge gap, $t^{*}_{\bot}\propto \Delta_{\rho}$.\cite{comment}
While in a simple two-chain system no real deconfinement is
observed,\cite{Tsuchiizu01} for a large (infinite) number of
chains\cite{Berthod06} the crossover is expected at some critical
value
$t^*_{\perp}\approx\Delta_{\rho}/2$.\cite{Suzumura98,Vescoli98}

If the temperature or energy scale at which the response of
the system is probed becomes larger than $t_\perp$, the warping of
the open Fermi surface is masked. Therefore, the properties of the
system are virtually the same as in the case of a flat Fermi
surface, i.e., for a strictly 1D system where the interacting
electrons are described by the Luttinger liquid (LL)
model.\cite{Schulz91} Thus, the deconfinement transition at high
enough temperature leads to a crossover from a Mott insulating to
a LL metallic state (see trajectories 2 and 4 in
Fig.~\ref{fig:deconfine}). Furthermore, a dimensional crossover
from a LL to a high-dimensional metallic state occurs when the
interchain coupling (i.e., the warping of the Fermi surface)
becomes larger than the energy of thermal fluctuations ($t_\perp >
k_B T$). This can be achieved by increasing the transfer
integral $t_{\perp}$ (trajectory 3) or by lowering the temperature
(trajectory 5). In the high-dimensional metallic state the
Fermi-liquid description is valid instead of the LL model.

Suitable systems to study the predicted dimensional crossover and
deconfinement transition are the well-known organic
Bechgaard-Fabre salts\cite{Jerome82,Ishiguro98,Jerome04}
(TMTTF)$_2X$ and (TMTSF)$_2X$. They consist of weakly coupled
molecular stacks along which the charge transport occurs. Hereby,
the type of the molecule (TMTSF or TMTTF) and of the anions $X^-$
determine the coupling between the molecular stacks and thus the
dimensionality of the system (chemical pressure effect). In
particular, the (TMTTF)$_2X$ salts consist of basically uncoupled
chains, which are half-filled due to the dimerization along the
chains. Thus, they constitute systems of electrons confined to the
chains, forming a Mott insulating state, and are located on the
left side of the diagram given in Fig.~\ref{fig:deconfine}.
At low temperatures the Mott insulating state of (TMTTF)$_2X$ salts is
transformed into a charge-ordered state characterized by a
disproportion of charge of the donor TMTTF molecules within the
stacks. This ferroelectric transition is strongly
pressure-dependent, and the charge-ordered phase can be suppressed
by the application of a moderate pressure of a few
kbars.\cite{chargeorder,Zamborszky02}
On the other hand, the compounds (TMTSF)$_2X$ represent
examples of quasi-1D metals and they are on the right side of the
diagram, since the enhanced interchain coupling leads to a charge
deconfinement.

The low-temperature optical conductivity spectra of the TMTSF
salts along the stacks are characterized by two distinct features:
a narrow mode at zero energy containing a very small part of the
spectral weight and a high-energy mode centered at around
200~\icm.\cite{Schwartz98} The latter feature was interpreted in
terms of the Mott (charge) gap of a 1D insulator.\cite{Favand96}
The frequency dependence of the optical conductivity in the energy
range above the effective interchain transfer integral [about
250~\icm\ for \pf] is well described by a LL model, since the
system can be considered as effectively one-dimensional. On the
other hand, the zero-energy mode was suggested to originate from
the effective doping of the 1D Mott insulator due to the
deconfined interchain electron hopping. Thus, it is related to 2D
physics and could be successfully described by the Fermi-liquid
model.\cite{Schwartz98} Similar on-chain optical conductivity
spectra with the dominating Mott gap were also observed for TMTTF
compounds.\cite{Vescoli99} However, the zero-energy mode typical
for TMTSF salts is absent, since these systems are in the confined
state where interchain hopping is suppressed. Thus, optical
spectroscopy on the high-energy range (larger than the interchain
transfer integral) addresses electronic interactions in 1D, while
dc resistivity measurements are more susceptible to interchain
hopping and thus reveal deviations from a strictly 1D picture.

First hints for the deconfinement transition and the dimensional
crossover were found by pressure-dependent transport measurements
on (TMTTF)$_2$PF$_6$ and \pf\ crystals.\cite{Moser98} The effect
of the dimensional crossover occurring during the temperature
decrease was addressed by an optical study of
\pf.\cite{Jacobsen81} Lateron, the above-mentioned theoretical
criterium for the deconfinement transition was verified by
comparing the ambient-pressure optical spectra of different
Bechgaard-Fabre salts,\cite{Vescoli98} i.e., making use of the
effect of chemical pressure on the interchain coupling. However,
this approach has disadvantages, since only a limited number of
discrete pressure points corresponding to particular compounds are
available. Moreover, atomic substitution (especially for sulfur
and selenium atoms) may appreciably affect not only the interchain
transfer integral $t_\perp$, but also the stack dimerization, on-chain
bandwidth, Coulomb repulsion etc.

A more direct and cleaner way to study the deconfinement
transition experimentally is the application of external pressure,
since the changes in the electrodynamic response can be monitored
while the interchain coupling is being continuously tuned. In this
paper we present the results of the first pressure-dependent infrared
spectroscopic study on the salt \asf\, which is a prime example of a
1D Mott-Hubbard insulator at ambient pressure.\cite{Dumm04,Zamborszky02,Moser98}
The main aim of the present investigation is to determine the
pressure-induced changes in the
optical response of \asf\ during the deconfinement transition and
the dimensional crossover. During our experiment we thus intended to
follow the evolution of the system along the dashed arrow in
Fig.~\ref{fig:deconfine}. A quantitative analysis of the measured
spectra was performed, in order to extract the value of the
interchain transfer integral $t_{\bot}$ as a function of pressure.
The obtained results are compared to the ambient-pressure optical
response of different Bechgaard-Fabre salts and discussed in
terms of the expected crossover from insulating (TMTTF)$_2X$ to
metallic (TMTSF)$_2X$ compounds.

\section{Experimental}\label{experimental}
Single crystals of \asf\ were grown by standard
electrochemical growth procedure. A diamond anvil cell equipped
with type IIA diamonds suitable for infrared measurements was used
to generate pressures up to 6~GPa. Finely ground CsI
powder was chosen as quasi-hydrostatic pressure transmitting
medium. The measured samples had as-grown specular surfaces
perpendicular to the $c$ axis and a thickness of about 50 $\mu$m.
The lateral dimensions were approximately $80\times
80$~$\mu$m$^2$. The pressure in the cell was measured using the
ruby luminescence method.\cite{Mao86} Polarized reflectivity
spectra were recorded at room temperature utilizing a Bruker IFS
66v/S FT-IR spectrometer in the mid-infrared (mid-IR) frequency
range (550 - 8000 \icm). To focus the beam on the small sample in
the pressure cell, an infrared microscope (Bruker IRscope II)
coupled to the spectrometer was used.

The intensity spectrum $I_{d}(\omega)$ reflected from the lower
air-diamond interface of the diamond anvil served as reference.
The obtained reflectivity spectra $R_{s-d}(\omega)$ refer to the
absolute reflectivity at the sample-diamond interface calculated
according to
\begin{equation}
  R_{s-d}(\omega)=R_{dia}\cdot\frac{I_{s-d}(\omega)}{I_{d}(\omega)},
  \label{eq:Rsd}
\end{equation}
\begin{figure}
  \includegraphics[width=1\columnwidth]{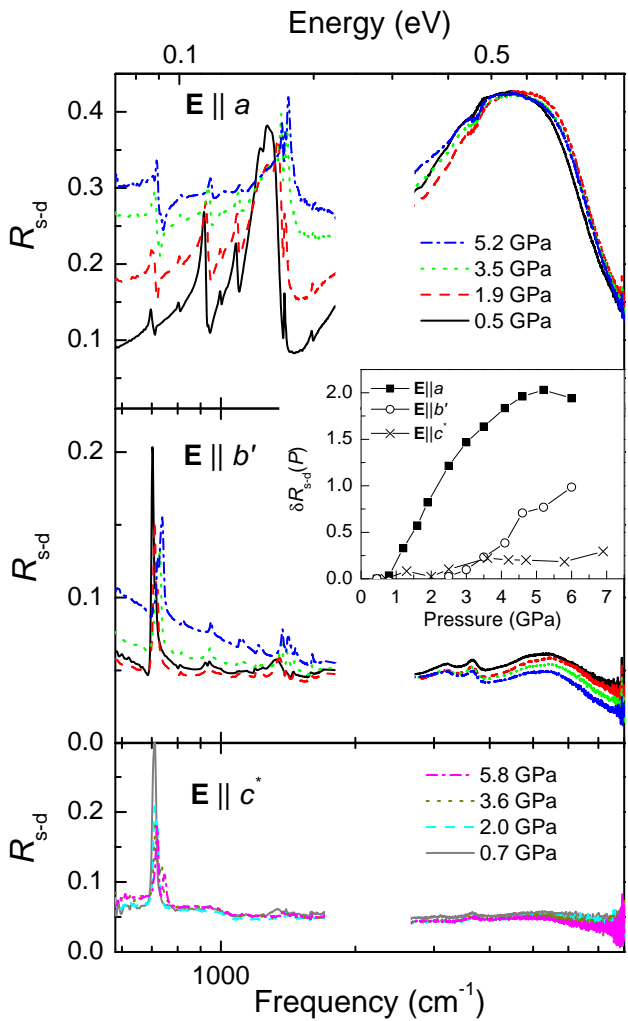}\\
  \caption{(Color online) Reflectivity spectra $R_{s-d}$ of \asf\ at room temperature
  as a function of pressure for \textbf{E}$\| a$, \textbf{E}$\| b'$, and
\textbf{E}$\| c^*$ at room temperature. Note the different
pressure values for \textbf{E}$\| a,b'$ and \textbf{E}$\| c^*$.
The inset shows the normalized pressure-induced change of the
reflectivity $\delta R_{s-d}(P)=R_{s-d}(P)/R_{s-d}$$(P_{\rm
min})$-1 at 630~\icm.\label{fig:refl}}
\end{figure}
where $I_{s-d}$ is the intensity reflected from the sample-diamond
interface and $R_{dia}$ is the absolute reflectivity of the
diamond-air interface. Since the refractive index of
diamond\cite{Palik} $n_d = 2.38$ is almost frequency independent
below $10^4$~\icm, $R_{dia}\simeq 0.167$ was used in our analysis.
Furthermore, it was assumed to be
pressure-independent.\cite{Eremets92} The reflectivity spectra
$R_{s-d}(\omega)$ in the frequency range 1700 - 2700~\icm\ are
affected by multiphonon absorptions of diamond, which cannot be
perfectly compensated. Therefore, this part of the measured
spectra is not shown.
In order to check the reproducibility of the results, we have
performed four high-pressure measurements on different \asf\
crystals.
\begin{figure}
  \includegraphics[width=1\columnwidth]{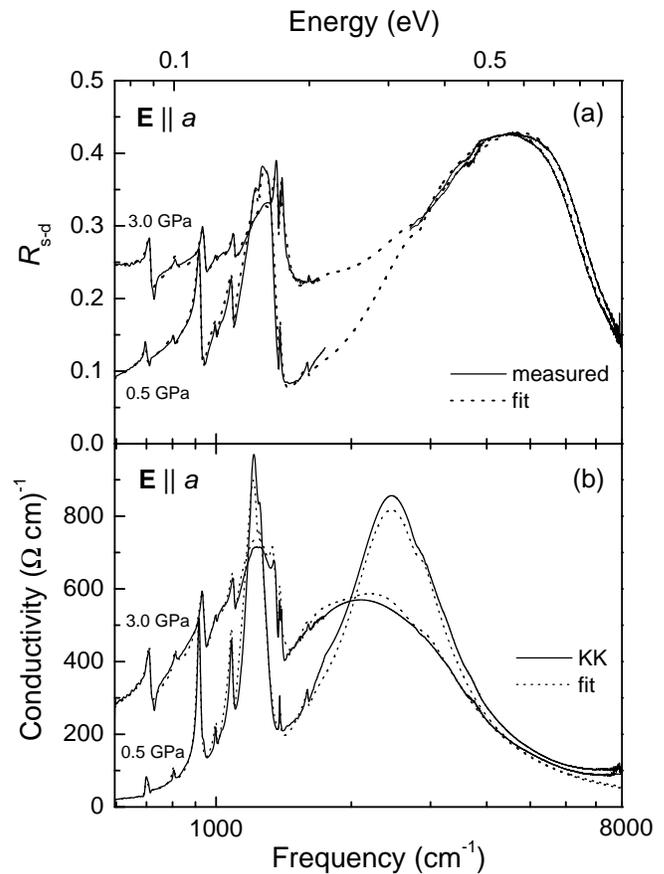}\\
  \caption{(a) Comparison of the measured reflectivity $R_{s-d}$
  of \asf\ with the fits, according to Eqs.\ (\ref{eq:4p-fit}) and
  (\ref{eq:ref-fit}),
  for \textbf{E}$\|a$ for two pressures (0.5 and 3.0~GPa);
  (b) corresponding optical conductivity $\sigma_1(\omega)$ spectra from the fit
  compared with the optical conductivity obtained by the KK analysis.\label{fig:fitting}}
\end{figure}
\section{Results}\label{results}
Pressure-dependent reflectivity spectra are presented in
Fig.~\ref{fig:refl} for the polarizations \textbf{E}$\| a$
(stacking direction), \textbf{E}$\| b'$ (direction perpendicular
to $a$, in the $a$-$b$ plane), and \textbf{E}$\| c^*$ (perpendicular
to $a$ and $b'$). Since the spectra for \textbf{E}$\| c^*$
were obtained in a separate experimental run, the pressure
values are slightly different from those for \textbf{E}$\| a,b'$.

The \textbf{E}$\|a$ reflectivity spectra include a strong mid-IR
band at around 4600~\icm\ and a number of sharp vibrational modes
below 2000~\icm. Most of them are totally symmetric
electron-molecular-vibration (emv) coupled A$_g$ Raman modes of
the TMTTF molecule.\cite{Jacobsen83} The application of
pressure leads to the gradual increase of the low-frequency part
($<$2000~\icm) of the reflectivity. At the same time, the
reflectivity level of the mid-IR band remains almost constant in
the whole studied pressure range.\par

For \textbf{E}$\| b'$ the reflectivity level is almost frequency-independent
at low pressures ($<$2~GPa), except for the
broad mid-IR absorption band located at around 5300~\icm. The
sharp peak at around 700~\icm\ is related to the F$_{1u}$ symmetry
vibration of the octahedral AsF$_6$ anion.\cite{Begun67} Other
weaker vibrational features for \textbf{E}$\| b'$ can be assigned
to the B$_{1u}$ and B$_{2u}$ IR-active modes of the TMTTF
molecules.\cite{Meneghetti84} The application of high enough
pressure ($>$2~GPa) results in a pronounced increase of the
reflectivity in the low-frequency range ($<$2000~\icm). At the
same time, the intensity of the mid-IR band decreases, while its
frequency position remains almost unchanged. In contrast to the
other studied directions, the reflectivity for {\bf E}$\parallel$$c^*$
is very low, almost frequency-independent (except for the sharp AsF$_6$
vibration at 700~\icm), and does not change with the application of
pressure up to 6 GPa.

The value of the low-frequency reflectivity ($<$2000~\icm) is
mostly affected by the Drude contribution due to free electrons.
Therefore, the pressure-induced enhancement of the charge
transport should lead to a higher reflectivity at low frequencies.
This is illustrated in the inset of Fig.~\ref{fig:refl}, where the
change of the reflectivity $\delta
R_{s-d}(P)=R_{s-d}(P)/R_{s-d}$$(P_{\rm min})$-1 at 630~\icm\
normalized to the reflectivity at the lowest applied pressure
$P_{\rm min}$ is depicted as a function of applied pressure $P$. The
enhancement of the reflectivity along the stacks (\textbf{E}$\|
a$) is as large as a factor of two at 5.2~GPa. For \textbf{E}$\|
b'$ the increase becomes noticeable only above 3~GPa, and reaches
one at the highes applied pressure. In contrast, the spectra of
\textbf{E}$\| c^*$ show a very small increase of only
$\approx$0.2 at 6~GPa, which cannot be analyzed quantitatively.

\section{Analysis and discussion}

\subsection{Longitudinal optical response}\label{a-direction}

In order to extract the optical conductivity for \textbf{E}$\| a$
from the measured pressure-dependent reflectivity spectra, we have
carried out a Kramers-Kronig (KK) analysis. The application of
this method to our data is not trivial due to two reasons: (i) The
measured reflectivity spectra cover a limited frequency range; for
the KK analysis they need to be extrapolated to lower and higher
frequencies and interpolated within the diamond absorption range
(1700-2700~\icm). (ii) The standard KK relation between the
reflectivity and phase needs to be corrected when it is applied to
the sample-diamond interface, $R_{s-d}$, and the necessary
correction term contains an \textit{a priori} unknown
parameter.\cite{Plaskett63}\par

As a first step in the analysis of the reflectivity data
$R_{s-d}(\omega)$, we performed an approximate fitting of the
reflectivity data. Most of the observed phonon resonances are
characterized by an asymmetric lineshape and cannot be fitted by a
simple three-parameter Lorentz model. We therefore used an
extended four-parameter factorized oscillator
model\cite{Gervais83b} for the fitting, which is frequently
applied for the analysis of IR phonon modes. The dielectric
function in this model is described by
\begin{equation}
  \epsilon(\omega)=\epsilon_{\infty}\prod_{j=1}^{n}\frac{\omega_{LO_j}^2-\omega^2+i\omega \gamma_{LO_j}}{\omega_{TO_j}^2-\omega^2+i\omega
  \gamma_{TO_j}},
  \label{eq:4p-fit}
\end{equation}
where $\omega_{TO_j}$ and $\omega_{LO_j}$ are transverse and
longitudinal frequencies of the $j$th polar mode, respectively,
$\gamma_{TO_j}$ and $\gamma_{LO_j}$ their respective damping
constants, and $\epsilon_{\infty}$ the optical dielectric constant
far above the resonance frequencies. The reflectivity is then
calculated according to
\begin{equation}
  R_{s-d}=\left|\frac{\sqrt{\epsilon(\omega)}-n_d}{\sqrt{\epsilon(\omega)}+n_d}\right|^2.
  \label{eq:ref-fit}
\end{equation}
With this procedure we were able to fit the measured reflectivity
spectra rather precisely. As examples, we show the results of the fits
together with the measured spectra for two pressures (0.5 and 3~GPa) in
Fig.~\ref{fig:fitting}(a).
\begin{figure}
  \includegraphics[width=1\columnwidth]{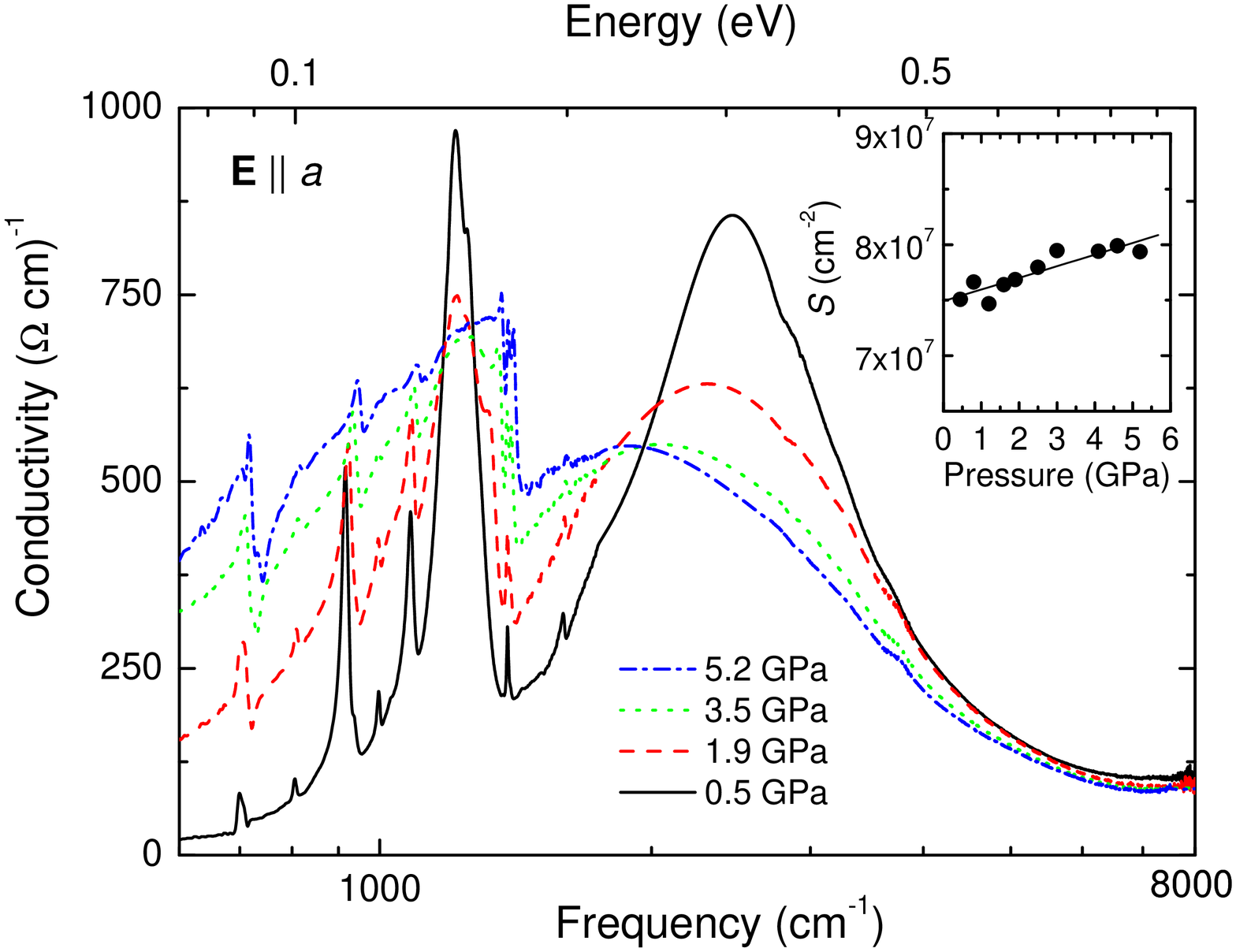}\\
  \caption{(Color online) Optical conductivity $\sigma_1(\omega)$ spectra of \asf\ for
  \textbf{E}$\| a$ as a function of pressure obtained by the KK analysis.
   Inset: Pressure dependence of the spectral weight $S$ according to Eq.~(\ref{eq:sumrule}).\label{fig:cond-a}}
\end{figure}

The obtained fits were used for the extrapolation of the
experimental spectra to lower and higher frequencies necessary for
the KK analysis. The reflectivity in the diamond absorption range
(1700 to 2700 \icm) was linearly interpolated. The KK
relation for the phase of the reflectivity $R_{s-d}$ has the
following form:\cite{Plaskett63,McDonald_phd}
\begin{equation}
  \phi(\omega_0)=-\frac{\omega_0}{\pi}P \int_{0}^{+\infty}\frac{\ln R_{s-d}(\omega)}{\omega^2-\omega_0^2}d\omega+\left[\pi-2\arctan\frac{\omega_\beta}{\omega_0}\right],
  \label{eq:KK}
\end{equation}
where $\omega_\beta$ is the position of the reflectivity pole on
the imaginary axis in the complex frequency plane. In case of
measurements on the sample-air interface, $\omega_\beta$ tends
towards infinity and the second term vanishes. For the
sample-diamond interface the second term must, however, be taken into
account. The criterium for the proper value of $\omega_\beta$ is
the agreement between the optical conductivity obtained by the KK
analysis and that from the initial fit. For the illustration of the applied
data analysis a comparison of these two spectra is given in Fig.~\ref{fig:fitting}(b)
for two different pressures; obviously, the agreement between the spectra is very
good. For all studied pressures the best correspondence between
the optical conductivity spectra obtained by fitting and the
output of the KK analysis is achieved for the parameter
$\omega_\beta = 4000$~\icm.

The pressure-dependent optical conductivity spectra
$\sigma_1(\omega)$ of \asf\ obtained by the KK analysis are
presented in Fig.~\ref{fig:cond-a}. For the lowest applied
pressure (0.5 GPa) the conductivity spectrum is very similar to
the published ambient-pressure data.\cite{Dumm04} It can be viewed
as a broad absorption band centered at around 2450~\icm\ with a
fine structure of vibronic modes.\cite{Jacobsen83} This mid-IR
band is usually attributed to the optical transitions across the
Mott-Hubbard gap.\cite{Favand96,Dressel96,Schwartz98} According to
our results, the application of pressure causes a gradual shift of
the position of the band to lower frequencies, accompanied by the
onset of a Drude-like conductivity.\par

The spectral weight $S$ of the absorption band was estimated by
integrating the optical conductivity $\sigma_1(\omega)$ up to
$\omega_c=7000$~\icm
\begin{equation}
 S=\int_{0}^{\omega_c} \sigma_1(\omega)d\omega \simeq
 \frac{\omega_{pa}^{2}}{8} = \frac{\pi N e^2}{2m_a}.
 \label{eq:sumrule}
\end{equation}
Here, $\omega_{pa}$ denotes the on-chain plasma frequency, $N$ the
electron density, and $m_a$ the electronic band mass in $a$
direction. The spectral weight obtained according to
Eq.~(\ref{eq:sumrule}) is almost pressure-independent (see inset
of Fig.~\ref{fig:cond-a}). The slight monotonic increase of the
spectral weight (about 5\% from the lowest to the highest
pressure) can be attributed to the increase of the charge density
due to the pressure-induced lattice compression.\cite{Gallois87}
\begin{figure}
  \includegraphics[width=1\columnwidth]{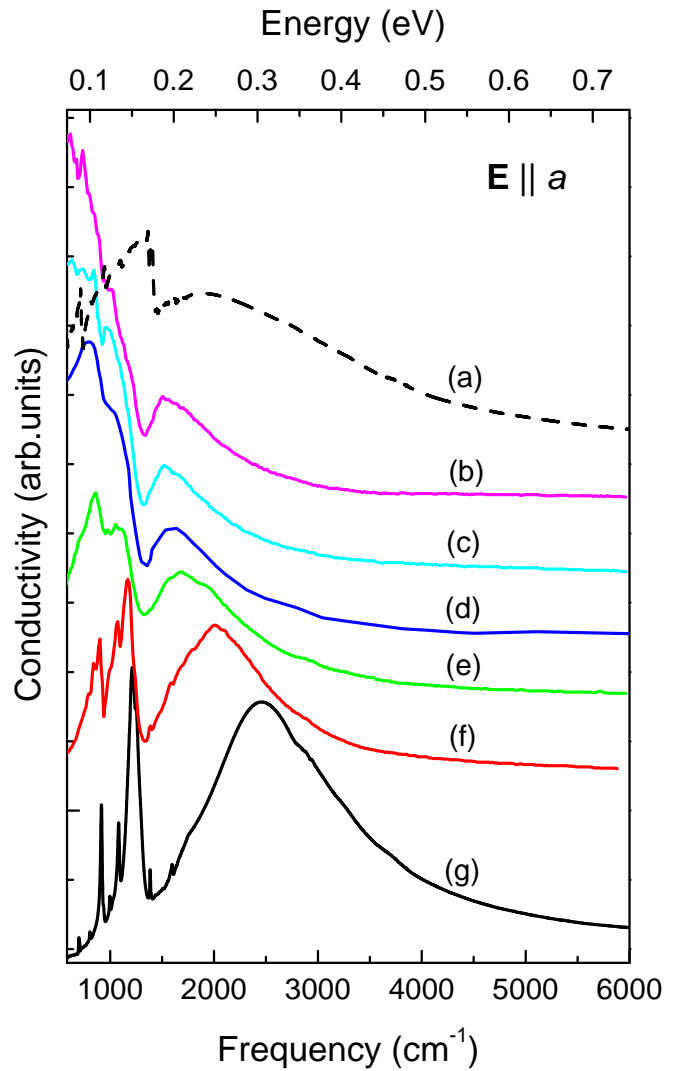}\\
  \caption{(Color online) Room temperature optical conductivity $\sigma_1(\omega)$ spectra
  of various Bechgaard-Fabre salts for \textbf{E}$\|
  a$: (a) \asf\ at 5.2 GPa; (b) (TMTSF)$_2$ClO$_4$; (c) (TMTSF)$_2$PF$_6$;
  (d) (TMTSF)$_2$AsF$_6$; (e) (TMTTF)$_2$Br;
   (f) (TMTTF)$_2$PF$_6$; (g) \asf\ at 0.5 GPa. The spectra
   (b)-(f) are taken from Refs.~\onlinecite{Pedron94,Schwartz98}.
   \label{fig:anion-compare}}
\end{figure}

Based on the spectral weight, we estimated the band mass $m_a$ of
\asf\ for the lowest pressure: with a charge density of
$N=1.5\times 10^{21}$~cm$^{-3}$, as calculated from the lattice
parameters,\cite{Laversanne84} we obtain a band mass
$m_a$$\approx$ $1.8 m_e$ ($m_e$ being the free electron mass)
according to Eq.~(\ref{eq:sumrule}). This value is larger than the
typical band mass in metallic TMTSF salts, where it is almost
equal to the free electron mass.\cite{Dressel96} The difference is
in good agreement with the calculated values of the room
temperature bandwidth for various Bechgaard-Fabre
salts.\cite{Ducasse86}

In Fig.~\ref{fig:anion-compare} the room-temperature
\textbf{E}$\|a$ optical conductivity spectra of various
Bechgaard-Fabre salts\cite{Pedron94,Schwartz98} are compared
with the corresponding results of the present \asf\ study at the
lowest and highest applied pressure. For clarity, the curves are
offset vertically and arranged from the bottom to the top with
increasing chemical pressure, according to the generic phase
diagram.\cite{Bourbonnais98} The spectrum of \asf\ at 5.2~GPa is
plotted above all other spectra. Obviously, the application of
\textit{chemical} pressure shifts the strong mid-IR band to lower
frequencies, suggesting the decrease of the Mott-Hubbard gap
energy which is mostly related to decrease of the stack
dimerization.\cite{Penc94,Favand96} As demonstrated in
Figs.~\ref{fig:cond-a} and \ref{fig:anion-compare}, the same
effects are also induced by the application of \textit{external}
pressure to \asf. However, a more detailed comparison reveals that
the above-mentioned effects induced by chemical pressure appear to
be less pronounced in the case of external pressure: even at a
relatively high pressure of 5.2~GPa the on-chain optical
conductivity of \asf\ differs from the ambient-pressure spectrum
of (TMTSF)$_2$AsF$_6$; it rather resembles that of (TMTTF)$_2$Br.
Thus, the spectral weight of the \textit{on-chain} optical
conductivity strongly depends on the type of cation (TMTSF or
TMTTF), while external pressure mainly changes the
\textit{inter-chain} coupling and the energy of the Mott-Hubbard
gap. This issue will be further discussed in
Section~\ref{b-direction}.

A measure of the strength of the electronic interaction can be obtained
by an analysis of the shape of the mir-IR band, i.e., the optical
transitions across the Mott-Hubbard gap.\cite{Dressel96,Schwartz98}
The optical conductivity well above the Mott-Hubbard gap of a 1D
insulator is expected to follow a power-law decay
according to\cite{Giamarchi91,Controzzi01,Giamarchi04}
\begin{equation}
 \sigma_1(\omega)\sim \omega^{- \gamma} = \omega^{4 z^2 K_{\rho} -
 5},
 \label{eq:power-law}
\end{equation}
where $z=2$ is the order of commensurability and $K_{\rho}$ is the
LL exponent characterizing the strength of electronic interaction.
$K_{\rho}=1$ corresponds to noninteracting electrons, $K_{\rho}<1$
to repulsive, and $K_{\rho}>1$ to attractive interaction between
the electrons. Based on the analysis of the low-temperature
optical conductivity of several TMTSF compounds, the power-law
exponent $\gamma$ was determined to be $\gamma \simeq 1.3$, which
corresponds to $K_{\rho} = 0.23$.\cite{Schwartz98} Since the
studied energy range is significantly higher than the thermal
energy ($\hbar \omega \gg k_B T$), Eq.~(\ref{eq:power-law}) should
be valid also for the room-temperature optical conductivity. An
analysis of the corresponding spectra in
Ref.~\onlinecite{Schwartz98} gives $\gamma$=$1.8 \dots 2.0$
($K_{\rho}=0.19 \dots 0.20$). This room-temperature value of
$K_\rho$ is only slightly smaller than the low-temperature value,
indicating the weak influence of temperature on the electronic
correlations in the Bechgaard-Fabre salts.
\begin{figure}
  \includegraphics[width=1\columnwidth]{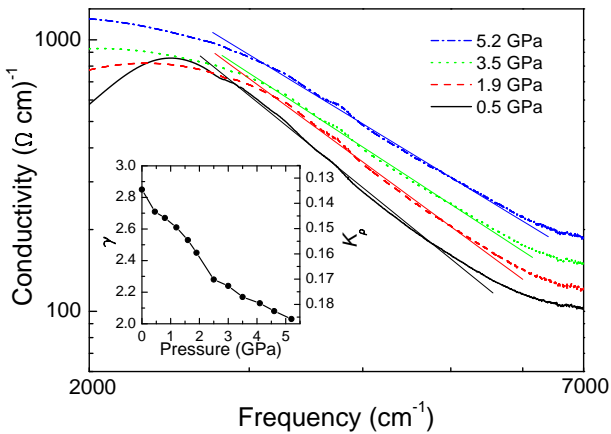}\\
  \caption{(Color online) The log-log scale plot of the optical
  conductivity spectra $\sigma_1(\omega)$ of \asf\ at different pressures. The straight lines
  are fits according to Eq.~(\ref{eq:power-law}).
  Inset: Power-law exponent $\gamma$ and corresponding LL exponent $K_\rho$ as a function of
  applied pressure.\label{fig:slopes}}
\end{figure}

We applied Eq.~(\ref{eq:power-law}) to the analysis of our optical
conductivity spectra of \asf. At ambient pressure we obtain
$\gamma=2.85$ ($K_\rho=0.13$), in agreement with the value
obtained by fitting the published spectra of (TMTTF)$_2$PF$_6$ and
(TMTTF)$_2$Br (3.0 and 2.5, resp.).\cite{Pedron94} This
demonstrates the reliability of our analysis. The pressure-induced
change of the power-law exponent is illustrated in
Fig.~\ref{fig:slopes}, where the measured conductivity spectra are
plotted in double-logarithmic scale. One can clearly observe a
gradual decrease of the slope with increase of pressure. The inset
of Fig.~\ref{fig:slopes} shows the power-law exponent $\gamma$ and
the respective value of $K_\rho$ as a function of pressure. At
5.2~GPa we obtain $\gamma\simeq 2$ and $K_\rho=0.19$ for \asf,
i.e., the typical value for TMTSF compounds. Obviously, the
application of external pressure leads to a gradual decrease of
the effective repulsive interaction strength in the studied
compound as it was generally claimed in the earlier
work.\cite{Schwartz98} Since $K_{\rho}$ is close to the value
$0.25$ at which quarter-filling umklapp scattering is irrelevant,
this means that
making the system less interacting decreases strongly the Mott gap
as it was observed in the previous study.\cite{Vescoli98} This
could explain the large variation of the Mott gap with pressure
assuming that the quarter-filled umklapp process dominates. Thus,
in principle, the effect of the half-filled umklapp due to the
stack dimerization is not necessary in order to explain the
electronic behavior in the Bechgaard-Fabre salts, although its
influence cannot be completely excluded.

Furthermore, the interchain hopping is enhanced as the distance
between the chains is reduced by external pressure. This effect
can be best followed in the transverse optical response as
discussed in the next Section.

\subsection{Transverse optical response}\label{b-direction}

In the reflectivity spectrum for \textbf{E}$\|c^*$, i.e., along the
direction perpendicular to $a$-$b$ plane, no pronounced electronic
excitations are found in the whole studied energy range. This indicates
an insulating behavior similar to the $c^{*}$ axis optical conductivity
in (TMTSF)$_2$ClO$_4$.\cite{Henderson99} The application of pressure up
to 5.8~GPa does not induce any appreciable change in the reflectivity
spectra.
\begin{table*}
\caption{\label{tab:bandwidth} Parameters obtained from the fit of
the \textbf{E}$\| b'$ reflectivity spectra of \asf\ with the Drude
model according to Eq.~(\ref{eq:Drude}): background dielectric
constant $\epsilon_\infty$, plasma frequency $\omega_p$, and dc
conductivity $\sigma_{dc}$=$\epsilon_0 \omega_p^{2} / \Gamma$.
Also listed are the auxiliary parameters (unit cell volume $V_c$
and carrier concentration $n$) and the interchain transfer
integral $t_b$ calculated according to Eq.~(\ref{eq:Kwak}).}
\begin{ruledtabular}
\begin{tabular}{ccccccc}
p (GPa) & $\epsilon_\infty$ & $\omega_p / 2\pi c$ (\icm)
& $\sigma_{dc}$ ($\Omega^{-1}$ cm$^{-1}$) & $V_c$ (\AA$^{3}$) & $n$ ($10^{21}$ cm$^{-3}$) & $t_b$ (meV)\\
\hline

2.5 & 2.80 & 2470 & 26.7 & 560.5 & 1.78 & 25\\

3.0 & 2.87 & 2710 & 32.1 & 541.1 & 1.85 & 27\\

3.5 & 2.96 & 3000 & 39.5 & 523.9 & 1.91 & 30\\

4.1 & 3.06 & 3300 & 47.6 & 506.1 & 1.98 & 33\\

4.6 & 3.15 & 3730 & 61.0 & 493.6 & 2.03 & 37\\

5.2 & 3.25 & 3840 & 64.5 & 481.3 & 2.08 & 38\\

6.0 & 3.33 & 4120 & 74.3 & 469.6 & 2.13 & 41\\

\end{tabular}
\end{ruledtabular}
\end{table*}
\begin{figure}[b]
  \includegraphics[width=1\columnwidth]{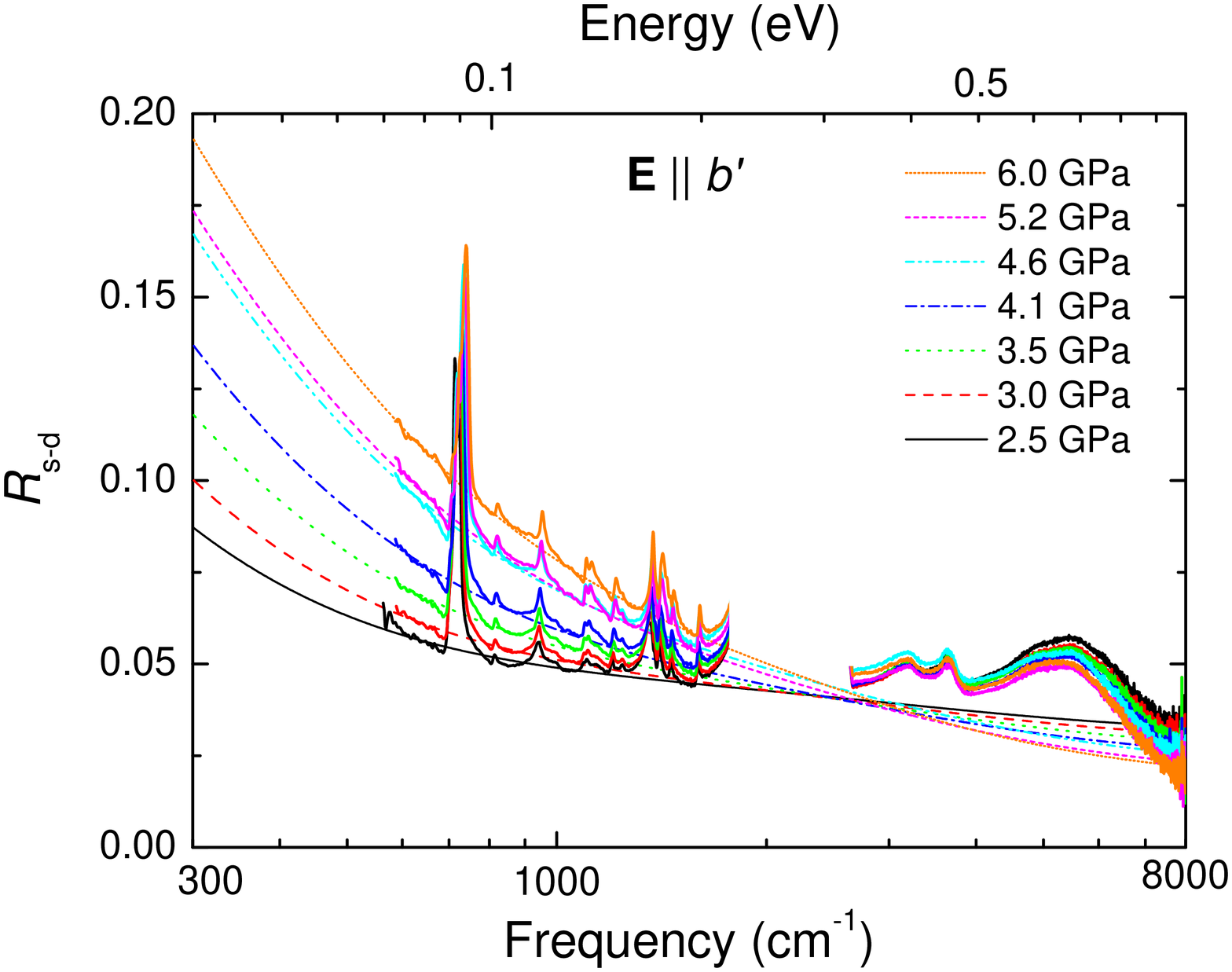}\\
  \caption{(Color online) Reflectivity spectra $R_{s-d}$ of \asf\ for
  \textbf{E}$\| b'$ together with the Drude fit according to Eq.~(\ref{eq:Drude}).\label{fig:drude-b}}
\end{figure}

The optical response along the $b'$ direction, i.e., perpendicular
to the stacks of TMTTF molecules, shows a pronounced
pressure-induced increase of the low-frequency reflectivity for
pressures above 2~GPa. This increase signals the onset of a
Drude-like contribution and thus the deconfinement of the charge
carriers. The Drude-like response can be described quantitatively
by estimating the transfer integral in $b'$ direction. For this
purpose, we fitted the low-frequency part of the measured
$R_{s-d}$ spectra for pressures above 2 GPa with a simple Drude
model, where the dielectric function is described by
\begin{equation}
  \epsilon(\omega) = \epsilon_\infty - \frac{\omega_p^2}{\omega (\omega +
  i\Gamma)},
  \label{eq:Drude}
\end{equation}
with $\epsilon_\infty$ being the background dielectric constant,
$\omega_p$ the plasma frequency, and the optical scattering rate $\Gamma$.
$\Gamma$ was considered to be
pressure-independent, since at room temperature it is mostly determined
by the scattering on thermal phonons. Its value $\Gamma \simeq
3800$~\icm\ agrees with literature data on
\pf.\cite{Jacobsen81,Jacobsen83} The measured reflectivity spectra of \asf\ together with
the fits are shown in Fig.~\ref{fig:drude-b}.
The parameters of the fits and the related dc conductivity,
$\sigma_{dc}=\epsilon_0 \omega_p^{2} / \Gamma$, where $\epsilon_0$
is the vacuum dielectric constant, are listed in
Table~\ref{tab:bandwidth}.

With the obtained values of $\omega_p$ we can calculate the
interchain transfer integral $t_b$ according to
\begin{equation}
  t_b^2 = \frac{\pi \epsilon_0 \hbar^2 V_c t_a \omega_p^2}{4 e^2 b^2},
  \label{eq:Kwak}
\end{equation}
valid for an open Fermi surface.\cite{Kwak82} Here $V_c$ denotes
the unit cell volume, $t_a$ the transfer integral along the
stacks, and $b$ the interstack separation. In a simple
tight-binding model $t_a$ should scale inversely proportional to
the effective mass. From the analysis of the spectral weight given
in Section~\ref{a-direction}, we found an effective mass $m_a=1.8
m_e$ which is almost pressure-independent. For Bechgaard salts
like \pf, which have an effective mass equal to the free electron
mass, $t_a$ is approximately 0.25~eV.\cite{Jacobsen83,Ducasse86}
Therefore, we estimated the transfer integral to be $t_a =
0.25/1.8 = 0.14$~eV for \asf\ at all pressures. Due to the lack of
high-pressure structural data on \asf, we had to make several
assumptions, in order to obtain $b$ and $V_c$ as a function of
pressure: The parameter $b$ was assumed to be linearly
pressure-dependent in the whole studied pressure range, with the
pressure coefficient 2.2~\%~GPa$^{-1}$ taken from the data of
Gallois et al.\cite{Gallois87} of \pf\ for pressures up to
1.6~GPa. The unit cell volume $V_c$ above 2~GPa was estimated from
the pressure-induced change of the background dielectric constant
$\epsilon_\infty(P)$ according to the Clausius-Mossotti
relation\cite{Ashcroft76b}
\begin{equation}
  \frac{\epsilon_\infty(P) - 1}{\epsilon_\infty(P) + 2} = \frac{\alpha_b}{3 \epsilon_0 V_c(P)},
  \label{eq:Clausius}
\end{equation}
where $\alpha_b$ is the electronic polarizability of the \asf\
unit cell in $b'$ direction. The value of $\alpha_b$ was chosen to
fit the known unit cell volume at pressures below
1.6~GPa\cite{Laversanne84,Gallois87} and was assumed to be
pressure-independent. The carrier density was calculated assuming
one carrier per unit cell. The resulting values of the unit cell
volume are listed in Table~\ref{tab:bandwidth}. We verified that
the variation of the lattice parameters affects the calculated charge
transfer integral only very weakly compared to the uncertainties
of the Drude fit.

\begin{figure}
  \includegraphics[width=1\columnwidth]{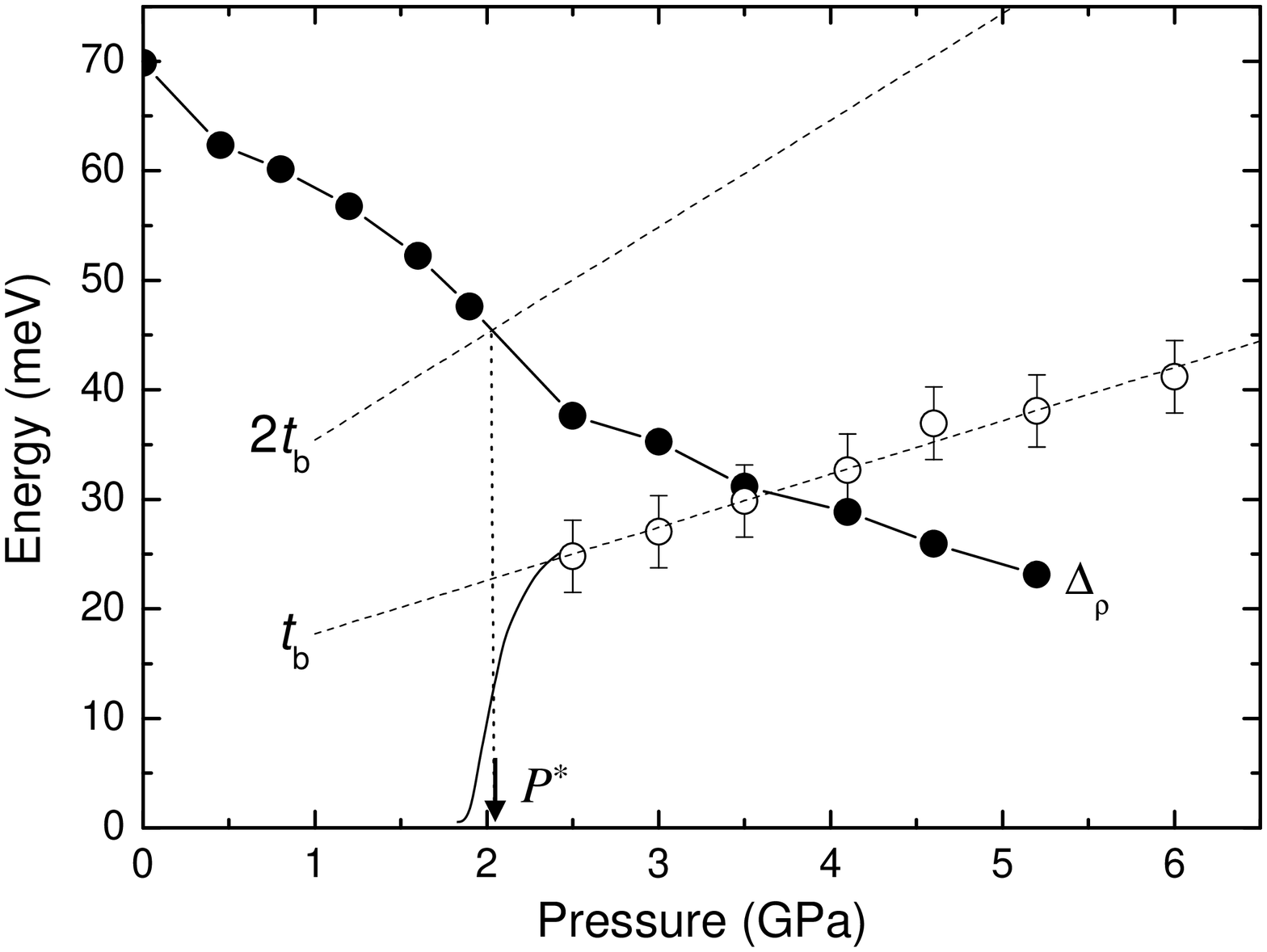}\\
  \caption{Calculated interchain transfer integral $t_b$ (open circles) and the charge gap $\Delta_{\rho}$ (full circles) as
  a function of pressure for \asf. The dashed lines are the linear fits of $t_b$ and
  $2t_b$. $P^*$ is the pressure at which the deconfinement transition takes place.
  The full line schematically shows the vanishing of $t_b$ at the transition pressure.
  \label{fig:bandwidth-b}}
\end{figure}

The pressure dependence of the transfer integral $t_b$
calculated according to Eq.~(\ref{eq:Kwak}) is shown in
Fig.~\ref{fig:bandwidth-b}. Above 2.0~GPa, $t_b$ increases
approximately linearly with increasing pressure, with
a linear pressure coefficient of $\approx$5~meV/GPa.
At a pressure of 4.5~GPa the value of the transfer integral in
\asf\ becomes equal to the ambient-pressure value in \pf\
($t_b=35$~meV).\cite{Jacobsen81,Kwak82} According to the generic
temperature-pressure phase diagram of the Bechgaard-Fabre
salts\cite{Bourbonnais98} the separation on the pressure scale
between \asf\ and \pf\ is about 3~GPa, i.e., appreciably smaller
than 4.5~GPa. Probably, the reason for this discrepancy is the way
in which the generic phase diagram was obtained. The various
Bechgaard-Fabre salts are arranged in the diagram mostly
according to their ground states (spin-Peierls, superconducting
etc.), whose nature is not determined solely by interchain coupling.
Therefore, one would not expect a perfect
agreement with our room temperature results related to the
equivalence of $t_b$ values.

With increasing pressure the deconfinement transition (trajectory
2 in Fig.~\ref{fig:deconfine}) is expected to occur when the
Mott-Hubbard gap $\Delta_{\rho}$ equals approximately
$2t_b$.\cite{Suzumura98,Vescoli98} In order to verify this
condition, we estimate the pressure dependence of the charge gap
$\Delta_\rho$ using the expression for a 1D quarter-filled
Mott-Hubbard insulator\cite{Schwartz98,Jerome04}
\begin{equation}
  2\Delta_{\rho}=W\left(\frac{g_{U}}{W}\right)^{1/(2-8K_{\rho})},
  \label{eq:gap}
\end{equation}
where $W\simeq 4t_a$ is the bandwidth and $g_{U}$ the coupling
constant of the corresponding umklapp process. The pressure
dependence of $K_\rho$ was obtained from the analysis of the
longitudinal optical response. As already discussed, our experiment
indicates that the bandwidth $W$ does not strongly depend on
pressure and amounts to 560~meV. Using the value of
$\Delta_{\rho}\simeq 70$~meV known from dc resistivity
measurements,\cite{Coulon82,Dressel01} the umklapp coupling
constant at ambient pressure can be calculated according to
Eq.~(\ref{eq:gap}) and amounts to $g_U=155$~meV. Since the increase
of $K_\rho$ with increasing pressure (see inset in
Fig.~\ref{fig:slopes}) indicates a decrease of the electronic
interaction, $g_U$ should decrease as well. The functional
relation between $g_U$ and $K_\rho$ is, however, not known for the
case of a strong non-local electron repulsion when $K_\rho < 0.5$.
Therefore, we calculated the pressure-dependent charge gap
$\Delta_\rho$ assuming that all parameters in Eq.~(\ref{eq:gap})
except $K_\rho$ are pressure-independent.

The resulting values of $\Delta_\rho$ as a function of pressure is
depicted in Fig.~\ref{fig:bandwidth-b}. Accordingly, for \asf\ the
deconfinement transition occurs at $P^*=2$~GPa. At the same time,
the interchain transfer integral $t_b$ related to single-particle
hopping should renormalize to zero in the confined
state\cite{Suzumura98,Tsuchiizu99} This is directly confirmed by
our experimental results: The Drude-like behavior is not observed
below 2~GPa indicating a strong suppression of the interchain
charge transfer, as illustrated by the full line in
Fig.~\ref{fig:bandwidth-b}.

According to the temperature-pressure phase diagram, which is
based on pressure-dependent resistivity measurements on the
closely related (TMTTF)$_2$PF$_6$ compound,\cite{Moser98} the
deconfinement transition occurs at a pressure of 1~GPa at room
temperature. This value is somewhat lower than the pressure of the
deconfinement transition in \asf\ obtained in the present study.
Possibly, this discrepancy is related to the criteria which were
used to specify the transition: In the mentioned work of Moser
\textit{et al.}\cite{Moser98} the deconfinement transition was
identified based on the resistivity behavior along the $a$ axis,
whereas in the present work we refer to the onset of the charge
transport along the $b'$ direction.

The dimensional crossover (trajectory 3 in
Fig.~\ref{fig:deconfine}) is expected to reveal itself as an onset
of a sharp plasma edge\cite{Jacobsen81} in the $b'$ direction,
i.e., the plasma frequency $\omega_p$ should become significantly
larger than the scattering rate $\Gamma$ ($\approx$3800 cm$^{-1}$).
According to the values of $\omega_p$ given
in Table~\ref{tab:bandwidth}, the Drude response is
overdamped (i.e., $\omega_p<2\Gamma$) in the whole studied
pressure range. Thus, even at 6~GPa \asf\ is presumably in a 1D LL
state, and the dimensional crossover is expected to occur only
at even higher pressure. According to the linear extrapolation
(not shown) of the pressure dependence of $\omega_p$ (which is
proportional to $t_b$) the condition $\omega_p=2\Gamma$ for the
dimensional crossover is fulfilled at a pressure of $P_{cr}$ = 12.8~GPa.
The estimated value of the interchain transfer integral at this
pressure is $t_b^{cr} \approx 75$~meV.

The so-obtained value of $P_{cr}$ does not agree with the results
of dc resistivity measurements along $c^*$ published by Moser et
al.\cite{Moser98} There, a strong increase of the dimensional
crossover temperature $T_{cr}$ with application of pressure was
demonstrated, such that $T_{cr}$ should reach room temperature at
3~GPa for (TMTTF)$_2$PF$_6$ and at 1~GPa for \pf. The former value
is smaller by a factor of about four compared to our estimation
based on the criterium of the underdamped Drude response.
Therefore, we suggest that this criterium should not be used for
the precise determination of the dimensional crossover pressure or
temperature. Rather, the observation of a well underdamped Drude
response ($\omega_p \gg 2\Gamma$) could be taken as a criterium
that the system can no longer be considered as one-dimensional.

\section{Conclusions}
In summary, we studied the mid-IR reflectivity of \asf\ at room
temperature with the incident radiation polarized along the $a$,
$b'$, and $c^*$ direction as a function of quasi-hydrostatic
pressure.

The pressure dependence of the optical conductivity along the
molecular stacks ({\bf E}$\parallel$$a$) demonstrates a gradual
onset of a Drude-like response accompanied by a decrease of the
charge gap with increasing pressure. We found the band mass along the stacks $m_b=1.8m_e$
to be weakly pressure-dependent. Assuming that the 1D Hubbard
model describes the properties of \asf\ correctly in the
high-energy limit (i.e., above 2500~\icm), where the optical
conductivity decays according to a power law $\sigma_1(\omega)\sim
\omega^{-\gamma}$, we found the exponent $\gamma$ to decrease from
2.9 at ambient pressure to 2.0 at 5.2~GPa. The corresponding LL
exponent $K_\rho$ increases from 0.13 to 0.19 indicating a
weakening of the electronic interaction.

Perpendicular to the stacks the onset of an overdamped Drude-like
optical conductivity is observed above 2.5~GPa for {\bf
E}$\parallel$$b'$. The analysis of the data suggests a
linear pressure-induced change of the interchain transfer integral
$t_b$. The absence of
a noticeable Drude component below $P^*\approx2$~GPa and its appearance
above this pressure is interpreted in terms of a pressure-induced
deconfinement transition at $P^*$. For this pressure value
the interchain transfer integral $t_b$ is approximately equal to
half the charge gap energy $\Delta_\rho$. For {\bf
E}$\parallel$$c^*$ \asf\ remains insulating up to the highest
applied pressure.

\section{Acknowledgments}

We thank G. Untereiner for crystal growth,  as well as
M. Dumm and T. Giamarchi for discussions and comments.
Financial support by the DFG (Emmy Noether program) is
acknowledged.


\begin{thebibliography}{45}
\bibitem{comment}
 In principle, an effective interchain hopping $t^{\rm eff}_{\bot}$ needs to
 be taken into account, since the interchain interaction leads to a renormalization.
\end{thebibliography}
\end{document}